\documentclass[12pt,a4paper]{article}
\usepackage{epsfig}
\def\GeV{{\rm\, GeV}}

\def\beq{\begin{equation}}
\def\eeq{\end{equation}}
\def\bea{\begin{eqnarray}}
\def\eea{\end{eqnarray}}

\def\({\left(}   
\def\){\right)}   

\def\eq#1{{Eq.~(\ref{#1})}}

\def\npb#1#2#3{    {\it Nucl. Phys. }{\bf B#1} (19#2) #3}
\def\plb#1#2#3{    {\it Phys. Lett. }{\bf B#1} (19#2) #3}
\def\prd#1#2#3{    {\it Phys. Rev. }{\bf D#1} (19#2) #3}

\def\prl#1#2#3{    {\it Phys. Rev. Lett. }{\bf #1} (19#2) #3}

\def\zpc#1#2#3{    {\it Z. Phys. }{\bf C#1} (19#2) #3}

\def\sjnp#1#2#3{   {\it Sov. J. Nucl. Phys. }{\bf #1} (19#2) #3}

\begin{document}

\title { {\LARGE\bf{SCREENING CORRECTIONS IN PHOTO }}\\[2ex]    
{\LARGE\bf{AND DIS PRODUCTION OF $J/\Psi$}} }
\author{
{\bf
E.~Gotsman\thanks{E-mail: gotsman@post.tau.ac.il}~$\,^{a}$,
\quad E. ~Ferreira \thanks{E-mail: erasmo@if.ufrj.br}~$\,^{b}$,
\quad  E.~Levin\thanks{E-mail: leving@post.tau.ac.il,
 elevin@quark.phy.bnl.gov}~$\, ^{a}$,
}\\%[5mm]
{\bf 
 U.~Maor\thanks{E-mail: maor@post.tau.ac.il}~$\,^{a}$
 \quad and \quad
E. Naftali\thanks{E-mail: erann@post.tau.ac.}~$\,^{a}$ 
}\\[10mm]
 {\it\normalsize $^a$HEP Department}\\
 {\it\normalsize School of Physics and Astronomy,}\\ 
 {\it\normalsize Raymond and Beverly Sackler Faculty of Exact Science,}\\
 {\it\normalsize Tel-Aviv University, Ramat Aviv, 69978, Israel}\\[0.5cm]
{\it\normalsize $^b$  Instituto de Fisica, Universidade Federal do 
Rio de Janeiro}\\
{\it \normalsize Rio de Janeiro RJ 21945-970, Brasil}
}
\maketitle
\thispagestyle{empty}

\begin{abstract}
\noindent
Photo and DIS production of $J/\Psi$ are investigated and compared 
with calculations based on pQCD in the LLA approximation
without and with screening corrections. The calculation includes
corrections induced by the real part of the production amplitude, the
skewed (off diagonal) gluon distribution function and the relativistic
Fermi motion within the charmonium system.
Our pQCD results are also compared with the predictions obtained from 
a Regge type two Pomeron model. 
Our results show that the screened pQCD model gives a better reproduction
of the data than the non screened model. 
However, the available data does not enable us to exclude
any of the three models we have examined. The predictions of these models, 
when extrapolated to both low and very high energies may provide a more
effective discrimination between the different parameterizations. 
\end{abstract}

\begin{flushright}
\vspace{-20cm}
{\bf TAUP 2650-2000}\\
\end{flushright}

\thispagestyle{empty}

\newpage
\setcounter{page}{1}

\section{Introduction}
The measurement of high energy photo and deep inelastic scattering (DIS)
exclusive production of $J/\Psi$ serves as an important testing ground for
checking the validity of pQCD in the limit of a relatively small hardness
and small $x$.  The hardness of this process is fixed by
$m_c^2 \simeq \frac{1}{4}M_{J/\Psi}^2$, 
consequently\cite{RYS}\cite{BROD}\cite{RRML} 
it is a suitable means of
investigating the applicability and possible need for a re-formulation of
hard pQCD when approaching the kinematic interface with
the less understood npQCD soft domain. Indeed, over the past few years we
have been witness to vigorous experimental, phenomenological and
theoretical investigations of this process, which supplement the detailed
analysis of $F_2(x,Q^2)$, the proton structure function, and its
logarithmic derivatives in the small $Q^2$ and small $x$ limits.\\
\centerline{}
In the following we present a detailed study of $J/\Psi$ 
photo and DIS exclusive production. 
Being relatively well measured, it  
may help to check the self consistency of, and discriminate between, 
relevant theoretical models .
We recall that a key ingredient for 
determining the parton distribution functions (pdf) is the pQCD analysis 
of inclusive DIS which utilizes the DGLAP evolution equations\cite{DGLAP}.
These pdfs are then used as input for the pQCD calculation of
exclusive DIS channels (such as $J/\Psi$ production), which are usually
executed in the color dipole approximation\cite{dipole}. We shall
mainly be interested in the correlated problems of determining the gluon 
saturation scale, the role of screening corrections (SC) and the
relative importance of the soft npQCD component 
at the kinematic edge where the hard pQCD is applicable.\\
\centerline{}
As we shall specify in the next section, a realistic calculation 
of $J/\Psi$ photo and DIS production depends on a few
corrections to the original pQCD estimate:\\
\centerline{}
1) $K_R^2$ accounting for the real part of the production
amplitude\cite{real}. 
The pQCD calculation of \eq{Q} below relates to the imaginary part only.\\
\centerline{}
2) $K_G^2$ denoting the correction resulting from the contributions of the  
off diagonal (skewed) gluon distributions\cite{offdiagonal}.\\
\centerline{} 
3) $K_F^2$ denoting the correction due to Fermi motion of the heavy
quark within the quarkonium system\cite{Fermi}. As such,
it accounts for the deviations from the simple static non relativistic 
estimate of the vector meson wave function.\\
\centerline{}
These corrections contribute significantly to the final 
(amplitude squared) estimate. Thus, in our investigation we shall 
assess possible ambiguities in our calculations which may be associated 
with the estimated corrections to the bare pQCD calculation. 
Our analysis follows the same lines as those of our recent 
paper\cite{GFLMN} on $\partial F_2/\partial \ln Q^2$.

\section{ Photo and DIS production of $J/\Psi$}
The procedure for calculating the  
forward differential cross section for photo and DIS production
of a heavy vector meson in the color dipole approximation
is straightforward.
The calculation is
performed\cite{RYS}\cite{BROD}\cite{RRML}\cite{FMS}\cite{MRT}
in the LLA, assuming the produced vector meson quarkonium system to be
non relativistic.
The contribution of  
pQCD to the imaginary part of the $t=0$ differential cross
section of photo and DIS production of heavy vector mesons 
is given by 
\beq\label{Q}
\(\frac{d\sigma(\gamma^* p \rightarrow V p)}{dt}\)_{t=0}^{pQCD}\,=\,
\frac{\pi^3\Gamma_{ee}M_V^3}{48\alpha}\,
\frac{\alpha_S^2({\bar Q}^2)}{{\bar Q}^8}\,
\(xG^{DGLAP}(x,{\bar Q}^2)\)^2 \,\(1\,+\,\frac{Q^2}{M_V^2}\),
\eeq
where, $xG^{DGLAP}$ is the gluon distribution function as obtained from
the DGLAP analysis. In the non relativistic limit we have
\begin{eqnarray}
&{\bar Q}^2\,=\,\frac{M_V^2\,+\,Q^2}{4},\nonumber\\
&x\,=\,\frac{4{\bar Q}^2}{W^2}.\label{Q1}
\end{eqnarray}
In the following we discuss the photo and DIS production of
$J/\Psi$ as there is an abundance of data available in this channel 
\cite{ZEUSJ}\cite{H1J}\cite{fixedJ} 
spanning a relatively wide energy range. From a theoretical point of
view, its hardness is comparable to those we
have investigated in our $\partial F_2/\partial \ln Q^2$ 
analysis\cite{GFLMN}. In this study we have shown that the recent HERA 
data on the $Q^2$ logarithmic slope of $F_2$ 
is well reproduced by DGLAP with either the CTEQ5HQ pdf input\cite{CTEQ}, 
or by the GRV98NLO input\cite{GRV} which is corrected 
for SC\cite{GLMN}.
As we wish to maintain the compatibility of the $F_2$ and the
$J/\Psi$ production interpretations, we have confined our pQCD 
calculations only to the above two pdfs.
In order to compare with the experimental data, which are given as     
integrated cross sections, to \eq{Q}, we need to know $B$ - the 
$J/\Psi$ forward differential cross section slope. 
The experimental values are approximately
constant with a possible moderate energy dependence\cite{ZEUSJ}.
Theoretically, each of the models we shall consider has a somewhat
different estimate of $B$ which we shall specify. \\
\centerline{}
The main difficulty with a pQCD analysis of $J/\Psi$ is
the observation that the simple dipole calculation needs 
to be corrected for the following reasons:\\
\centerline{}
1) A correction  for the contribution of the real part of the 
production amplitude. 
This correction is well understood\cite{real} and is given by
\begin{eqnarray}
&K_R^2 \,= \,(1\,+\rho^2),\nonumber\\
&\rho \,= \, ReA/ImA\,=\,tg(\frac{\pi \lambda}{2}),\nonumber\\
&\lambda \,= \,\partial \ln (xG^{DGLAP})/\partial \ln
(\frac{1}{x}).\label{RR}
\end{eqnarray} 
We note that $\lambda$ has a mild dependence on $x$ at a fixed
${\bar Q}^2$.\\
\centerline{}
2) A correction  for the contribution of the skewed (off
diagonal) gluon distributions. This correction is
calculated\cite{offdiagonal} to be 
\beq
K_G^2\,=\,
\(\frac{2^{2\lambda+3}\,\Gamma(\lambda+2.5)}
{\sqrt{\pi}\,\Gamma(\lambda+4)}\)^2.\label{RG}
\eeq
3) A more controversial issue relates to the non relativistic
approximation assumed for the $J/\Psi$ charmonium. Relativistic effects,
produced by the Fermi motion of the bound quarks, result in a considerable
reduction of the calculated pQCD cross section\cite{Fermi}. 
The correction, $K_F^2$, is very sensitive to
the value of $m_c$. Ref.\cite{Fermi} assumes that $m_c \simeq 1.50\, GeV$
and obtains $K_F^2\simeq 0.25$ with almost no energy dependence. 
However, we note that the calculation of $K_F^2$, regardless of its
detailed construction, is very sensitive to the c-quark mass  
since a small change in the input value of $m_c$
changes the estimate of $K_F^2$ significantly. Clearly, if we assume that 
$m_c=\frac{1}{2}M_{J/\Psi} \simeq 1.55\, GeV$, i.e. a change of only about
$50\, MeV$ relative to the value assumed in Ref.\cite{Fermi}, the
Fermi correcting factor is identical to 1.
We will, therefore, consider $K_F^2$ as a free (energy
independent) parameter. 
As we shall see, a calculation based on CTEQ5 gives $K_F^2 \simeq 1.00$,  
whereas a calculation based on GRV98 with SC gives 
$K_F^2=0.70$. This value corresponds to a c-quark mass of approximately 
$1.52\, GeV$.\\
\centerline{}
Based on the above, the expression for the integrated cross section
is written
\beq\label{NSC}
\sigma(\gamma^* p \rightarrow J/\Psi p)\,=\,
K_R^2 \cdot K_G^2 \cdot K_F^2 \cdot \frac{1}{B}\cdot
\(\frac{d\sigma}{dt}\)_{t=0}^{pQCD},
\eeq
where $\(\frac{d\sigma}{dt}\)_{t=0}^{pQCD}$ is given by \eq{Q}, B (the
forward differential slope) is taken from the data and $K_F^2$ is a free
parameter.\\
\centerline{}
We wish to present an analysis for $J/\Psi$ photo and DIS pQCD cross
sections which is compatible with our recent investigation\cite{GFLMN} 
of the $Q^2$ logarithmic slope of $F_2$.
Note that the
$F_2$ DGLAP analysis has been performed in NLO while the present
calculation is carried out in the LLA modified by the
above corrections.
In the $F_2$ investigation we have shown\cite{GFLMN} that 
the recent logarithmic slope data is well
reproduced by three, rather different, formulations:\\
\centerline{}
1) A DGLAP calculation with CTEQ5 pdf input. This calculation has no
explicit soft contribution.\\
\centerline{}
2) A DGLAP calculation with GRV98 pdf input modified by SC which are 
calculated\cite{GLMN} in the DLA. This
calculation, as well, has no explicit soft contribution.\\
\centerline{}
3) A Regge type two Pomeron parameterization\cite{DL2P} in which the hard
contribution is provided by a hard Pomeron, whereas the soft
contribution is presented by a soft Pomeron.
The two trajectories are parametrized to be
\begin{eqnarray}
&\alpha^H(t)\,=\,1.44+\,0.1t\nonumber\\
&\alpha^S(t)\,=\,1.08+\,0.25t
\end{eqnarray}
\\
\centerline{}
In the following we check and compare the above models with
the $J/\Psi$ cross sections.
Our data base has 49 photo production 
points, out of which 35 points are HERA data. We have also studied just
the 30 points at the high energy end with $W>50\,GeV$. 
The DIS data has 47 data points, all of which come from HERA and have
$W>50\GeV$. \\
\centerline{}
1) Our non screened pQCD calculation, denoted CTEQ5NSC,
is based on \eq{NSC} with CTEQ5 input 
for $xG^{DGLAP}(x,{\bar Q}^2)$. Our calculations for photo production,
compared
with the experimental data, are presented in Fig.1.
This calculation has only a hard sector and as such  
in our calculations we took a fixed $B=4.73\, GeV^{-2}$\cite{H1J}. 
We have adjusted the free parameter $K_F^2$ 
so as to fit the data base of the    
integrated $J/\Psi$ photo production and DIS cross section points.
The best fit has no Fermi suppression, i.e. $K_F^2=1.00$, and has the
following values for the corresponding $\frac{\chi^2}{ndf}$: \\
a) For the complete data base $\frac{\chi^2}{ndf}=2.12$. \\
b) For the photo production data $\frac{\chi^2}{ndf}=3.03$. \\
c) For the HERA photo production data $\frac{\chi^2}{ndf}=1.14$. \\
d) For high energy
photo production ($W>50\,GeV$) $\frac{\chi^2}{ndf}=0.92$. \\
\centerline{}
As seen in Fig.1, CTEQ5NSC overestimates the low energy data. 
In general, the overall
$x$ dependence of CTEQNSC is somewhat softer than the harder $x$
behaviour suggested by the data. 
We conclude that CTEQ5NSC does not produce a good fit to the complete data
base. However, its $\frac{\chi^2}{ndf}$ 
is considerably improved once we ignore
the low energy data. \\
\centerline{}
2) For the SC calculation, done in the DLA,
we follow our earlier publications\cite{GLMV} and define the
damping factors due to the screening in the quark
sector i.e. the percolation of the $c \bar c$ through the target.
This is given by the following expressions for the longitudinal and
transverse damping factors
%%%%%%%%%%%%%%%%%%%%%%%%%%%%%%%%%%%%%%%%%%%%%%%%%%%%%%%%%%%%%%%%%%%
\begin{figure}
\centerline{\epsfig{file=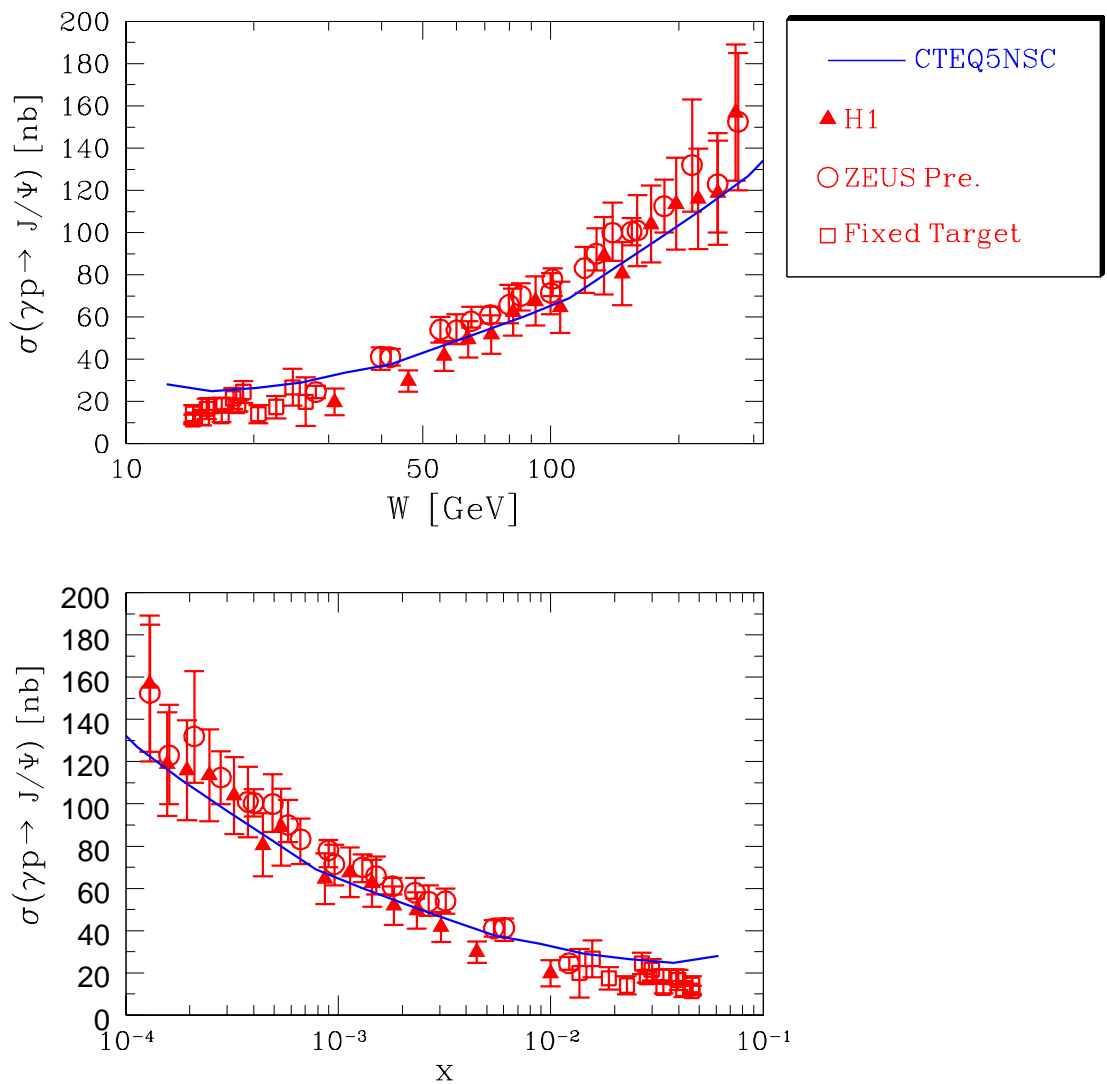,width=170mm}}
\caption{\it Photo production of $J/\Psi$ as a function of $W$ and
$x$. Data and CTEQ5NSC calculations.}
\label{Fig.1}
\end{figure}
%%%%%%%%%%%%%%%%%%%%%%%%%%%%%%%%%%%%%%%%%%%%%%%%%%%%%%%%%%%%%%%%%%%%
\beq\label{SCL}
D_{qL}^2\,=\,\(\frac{E_1(\frac{1}{\kappa_q})e^{\frac{1}{\kappa_q}}}
{\kappa_q}\)^2
\eeq
and 
\beq\label{SCT}
D_{qT}^2\,=\,
\left(\frac{1\,+\,(1\,-\,\frac{1}{\kappa_q})
E_1(\frac{1}{\kappa_q})e^{\frac{1}{\kappa_q}}}   
{2\kappa_q}\right)^2,
\eeq
where, for $N_c=3$, we have 
\beq
\kappa_q\,=\,\frac{2 \pi \alpha_S}{3 R^2 {\bar Q}^2}
xG^{DGLAP}(x,{\bar Q}^2).
\eeq
The above expressions are derived\cite{GLMV} assuming that 
$\gamma << 1$, where $\gamma$ is the DGLAP anomalous dimension.
As a result of this approximation we overestimate the SC in the
exceedingly small $x$ limit. However, within the kinematical range of this 
investigation this excess is small enough to be
neglected for $x>10^{-3}$. The correction for smaller $x$ may be as large
as $10\%$.
Actually, the exact calculation makes our results moderately 
harder and consequently improves our reproduction of the data (see Fig.2).
We do not consider this to be important as it will result in a small 
reduction of our present $\frac{\chi^2}{ndf}$ values, which 
are excellent as they stand.
Our expression for $D_g^2$, the damping in the gluon sector, is the square
of the gluon damping used in Refs.\cite{GFLMN}\cite{GLMN}.
Our final expression for the integrated cross section is
\beq\label{SC}
\sigma(\gamma^* p \rightarrow J/\Psi p)\,=\,
K_R^2 \cdot K_G^2 \cdot K_F^2 \cdot \frac{1}{B(R^2)} \cdot  
\(\frac{d\sigma}{dt}\)_{t=0}^{pQCD} \cdot D_q^2 \cdot D_g^2,
\eeq
where $D_q$ denotes the L and T components as appropriate. As in our $F_2$
study\cite{GFLMN}\cite{GLMN}, for  
$xG^{DGLAP}(x,{\bar Q}^2)$ we use the GRV98 pdf input. 
%%%%%%%%%%%%%%%%%%%%%%%%%%%%%%%%%%%%%%%%%%%%%%%%%%%%%%%%%%
\begin{figure}
\centerline{\epsfig{file=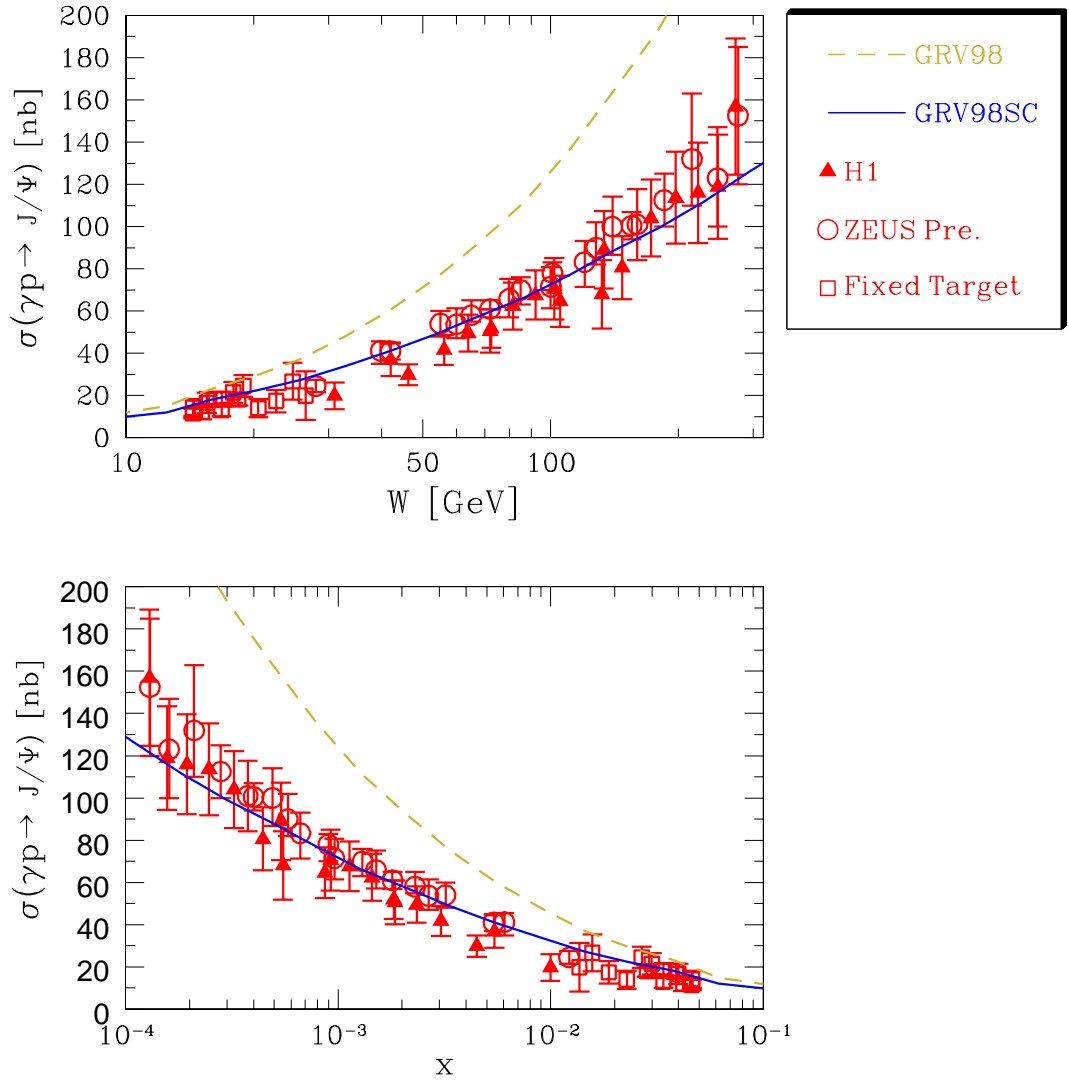,width=170mm}}
\caption{\it Photo production of $J/\Psi$ as a function of $W$ and
$x$. Data and GRV98 calculations with and without SC.}
\label{Fig.2}
\end{figure} 
%%%%%%%%%%%%%%%%%%%%%%%%%%%%%%%%%%%%%%%%%%%%%%%%%%%%%%%%%%%
\begin{figure}
\centerline{\epsfig{file=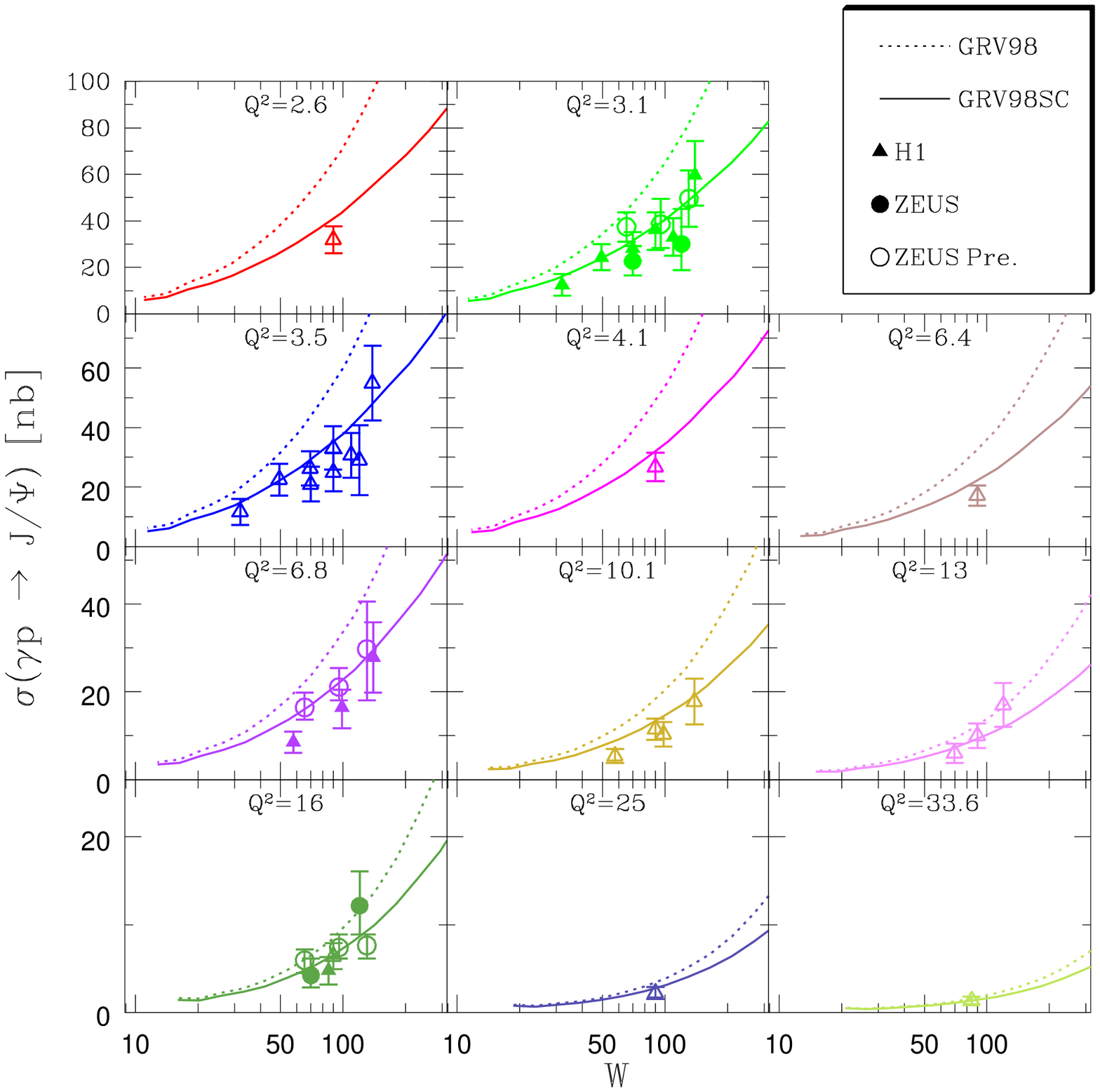,width=170mm}}
\caption{\it DIS production of $J/\Psi$. Data and our calculations.}
\label{Fig.3}
\end{figure} 
%%%%%%%%%%%%%%%%%%%%%%%%%%%%%%%%%%%%%%%%%%%%%%%%%%%%%%%%%%%%
\\
\centerline{}
Our calculations as compared with the data are presented in Figs.2 and 3.
with an adjusted value of $K_F^2=0.70$. We get the following values for
the corresponding $\frac{\chi^2}{ndf}$: \\
a) For the complete data base $\frac{\chi^2}{ndf}=0.94$. \\
b) For the photo production data $\frac{\chi^2}{ndf}=0.91$. \\
c) For the HERA photo production data $\frac{\chi^2}{ndf}=0.87$. \\
d) For high energy
photo production ($W>50\,GeV$) $\frac{\chi^2}{ndf}=0.58$. \\
\centerline{}
As is visible from Fig.2, the suppression induced by the SC is
appreciable even though $\kappa_g < 1$, i.e. below gluon saturation. 
This is consistent with the general observation\cite{Amirim} that SC,
which are the consequence of high gluon density, precede the gluon
saturation state. 
Note that $R^2=8.5\, GeV^{-2}$,
which is the essential parameter in the SC calculation presented both in 
Ref.\cite{GFLMN} and here, is not a free parameter but is determined   
directly from the $J/\Psi$ photo production forward slope. In a SC model,
such as ours, we expect\cite{GLMV2} a weak dependence of $B$ on x. 
This is demonstrated in Fig.4 together 
with the relevant HERA data. For a non screened calculation $B$ is 
fixed at $\frac{1}{2}R^2$. \\
\centerline{}
3) Donnachie and Landshoff (DL) have followed their $F_2$ 
analysis\cite{DL2P} with an updated 
publication on charm production\cite{DLcharm} in which they show that
the $J/\Psi$ photo and DIS data on the integrated and differential  
cross sections are compatible with their model, which we denote DL2P. 
We discuss the differences between DL2P and the predictions of 
CTEQ5NSC and GRV98SC in the next section.
%%%%%%%%%%%%%%%%%%%%%%%%%%%%%%%%%%%%%%%%%%%%%%%%%%%%%%%%%%%%%%%%%%%%
\begin{figure}
\centerline{\epsfig{file=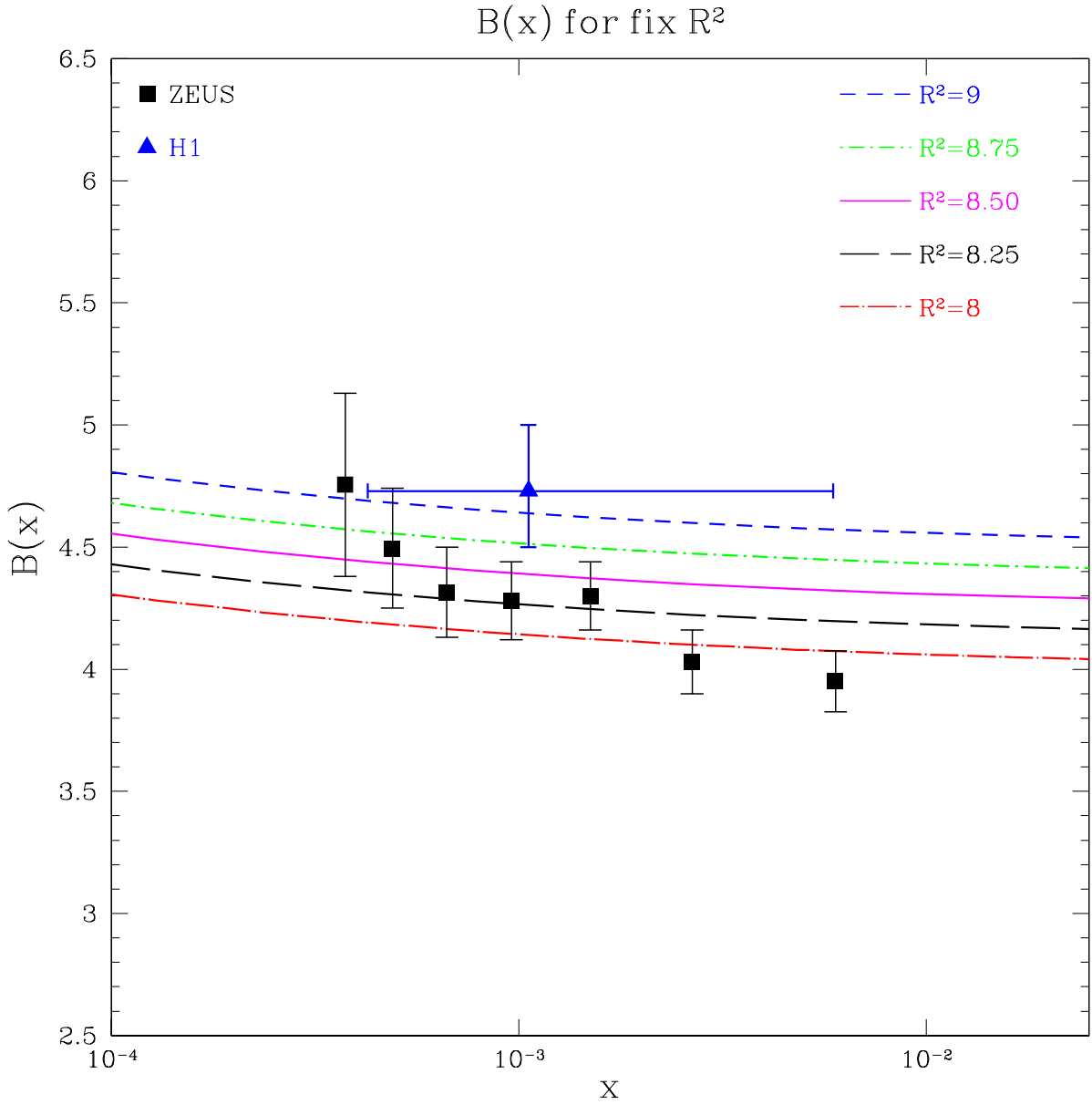,width=170mm}}
\caption{\it The energy dependence of the forward differential slope of
$J/\Psi$ photo production.
Data and SC calculations with several values
of $R^2$.}
\label{Fig.4}
\end{figure} 
%%%%%%%%%%%%%%%%%%%%%%%%%%%%%%%%%%%%%%%%%%%%%%%%%%%%%%%%%%%%%%%%%%%%

\section{Discussion and Conclusions}
The results of the present investigation corroborate the conclusions
obtained in our previous study\cite{GFLMN}
of $\partial F_2/\partial \ln Q^2$, where we have shown that the 
available data is consistent with either CTEQ5NSC, GRV98SC or DL2P 
in the kinematic range of $Q^2 \geq 1.9\, GeV^2$ and $x<10^{-2}$.
A comparison of the photo production (${\bar Q}^2=2.4\, GeV^2$)
predictions of these models over a wide $x$ range is presented in Fig.5. 
Following are some concluding remarks\\
\centerline{}
1) The differences between CTEQ5NSC and GRV98SC, within the
available experimental kinematic window, are relatively 
small. We note, though, a systematic difference at low energies
($x>10^{-2}$) where CTEQ5NSC is considerably larger than GRV98SC. This
difference is responsible for the high 
$\frac{\chi^2}{ndf}$ values we found for CTEQ5NSC. Similar differences
also exist for the $Q^2$ logarithmic slope predictions of these 
models\cite{GFLMN} in the low energy (high $x$) limit, but relevant data
is not available. 
From the study of $J/\Psi$ photo production we conclude that the  
GRV98SC parameterization of this process extrapolates well to the very low
energy domain in which the soft (npQCD) sector plays an increasingly
important role. On the other hand, CTEQ5NSC does not have
this property and it overestimates the low energy data.\\
\centerline{}
2) We also note a difference between CTEQ5NSC and GRV98SC in the high
energy region ($x<10^{-4}$).
This was not observed in our predictions for the logarithmic $F_2$ slope
and is attributed to our approximate calculation of $D_q^2(J/\Psi)$ which
results in a small excess of the SC for very small $x$. As we noted 
earlier, an exact calculation of $D_q^2(J/\Psi)$ will improve the results
of GRV98SC and bring it close to those of CTEQ5NSC. \\
\centerline{}
3) As we have shown, CTEQ5NSC provides an acceptable reproduction of the
high energy photo production data. These results are obtained with
$K_F^2=1.00$ which implies that the $J/\Psi$ charmonium bound system  
is strictly non relativistic, or else, that the normalization of
the calculation requires some adjustment.
We have further investigated this model by applying SC to the calculation 
with $K_F=1.00$, denoting it CTEQ5SC. Our output is non
satisfactory as it systematically underestimates the data.
This result supports our
suggestion\cite{GFLMN} that CTEQ5 may contain significant screening
effects
which are absent in the boundary conditions used in GRV98. In this 
context, it is suggestive to assume that 
CTEQ5SC results are too small due to a possible double counting of the
SC. Our calculated SC introduce a $20\%$ deficiency at $x=10^{-4}$.
We consider this as the uncertainty in the CTEQ5 parameterization in 
the exceedingly small $x$ limit. This 
is compatible with the similar conclusion reached in Ref.\cite{GFLMN}.  
The pQCD calculation
used in this investigation is performed in the LLA and is, thus, 
subject to corrections. We distinguish between the
normalization of the models we have discussed, which may change, and their
hardness as reflected in the energy, or $x$, dependence, which is a
more stable property.\\
\centerline{}
4) Within the available kinematic photo production window, DL2P predictions
which differ somewhat from the results of the two  
pQCD models, still provide an adequate reproduction of the 
data\cite{DLcharm}. At this stage, with the given experimental error bars,
one cannot conclusively exclude any of these models. At higher energies
DL2P is 
significantly larger than either CTEQ5NSC or GRV98SC. This feature of DL2P
is a consequence of the hard Pomeron component which has a very high Regge
trajectory intercept. As a result its energy
dependence is much harder than the pQCD models whose 
$\lambda$ values are considerably smaller. This is clearly seen 
at $x=10^{-5}$ where the DL prediction is about 5 times
larger than that of GRV98SC. This $x$ range will become
accessible at THERA and, thus, enable an experimental discrimination 
between DL2P and the pQCD models. The same behaviour of DL2P has also been
observed\cite{GFLMN} in the small $x$ limit of 
$\partial F_2/\partial \ln Q^2$. \\
\centerline{}
5) We have investigated the possible role of a soft component, 
parameterized in the DL form, which was added to GRV98SC. 
Our analysis shows that such an addition does not improve the 
$\frac{\chi^2}{ndf}$. We, thus, conclude that any soft component in our
model is exceedingly small. \\
\centerline{}
6) The three models discussed have different predictions for $B$, the
$J/\Psi$ forward differential slope. CTEQ5NSC is a model with only a
hard, non modified, sector. As such, we expect its $B$ to be a constant
for which we took the experimental value. GRV98SC is also a model with
only a hard sector but it is corrected by SC. As a result, the model
predicts a modest shrinkage of $B$ which is presented together with
the data in Fig.4. DL2P is a sum of a soft component, which dominates in
the low energy limit and a hard component which dominates in the high
energy limit. Since the two Pomerons have different trajectory slopes,
DL2P predicts an anti shrinkage effect, where $B$ decreases
monotonically with W. The recent ZEUS\cite{ZEUSJ} results 
contradict this feature. \\
\centerline{}
Overall, our investigation shows that   
GRV98SC results in a better reproduction of the data than the other two
models considered. However, the available data does not enable us, as yet,
to conclusively exclude any of the models we have examined.\\
\centerline{}
{\bf Acknowledgments:}
UM wishes to thank UFRJ and FAPERJ (Brazil) for their support.
This research was supported by in part by the Israel Science Foundation,
by BSF grant \# 98000276 and by GIF grant \#I-620-22.14/1999.
%%%%%%%%%%%%%%%%%%%%%%%%%%%%%%%%%%%%%%%%%%%%%%%%%%%%%%%%%%%%%%%%%%%%%
\begin{figure}
\centerline{\epsfig{file=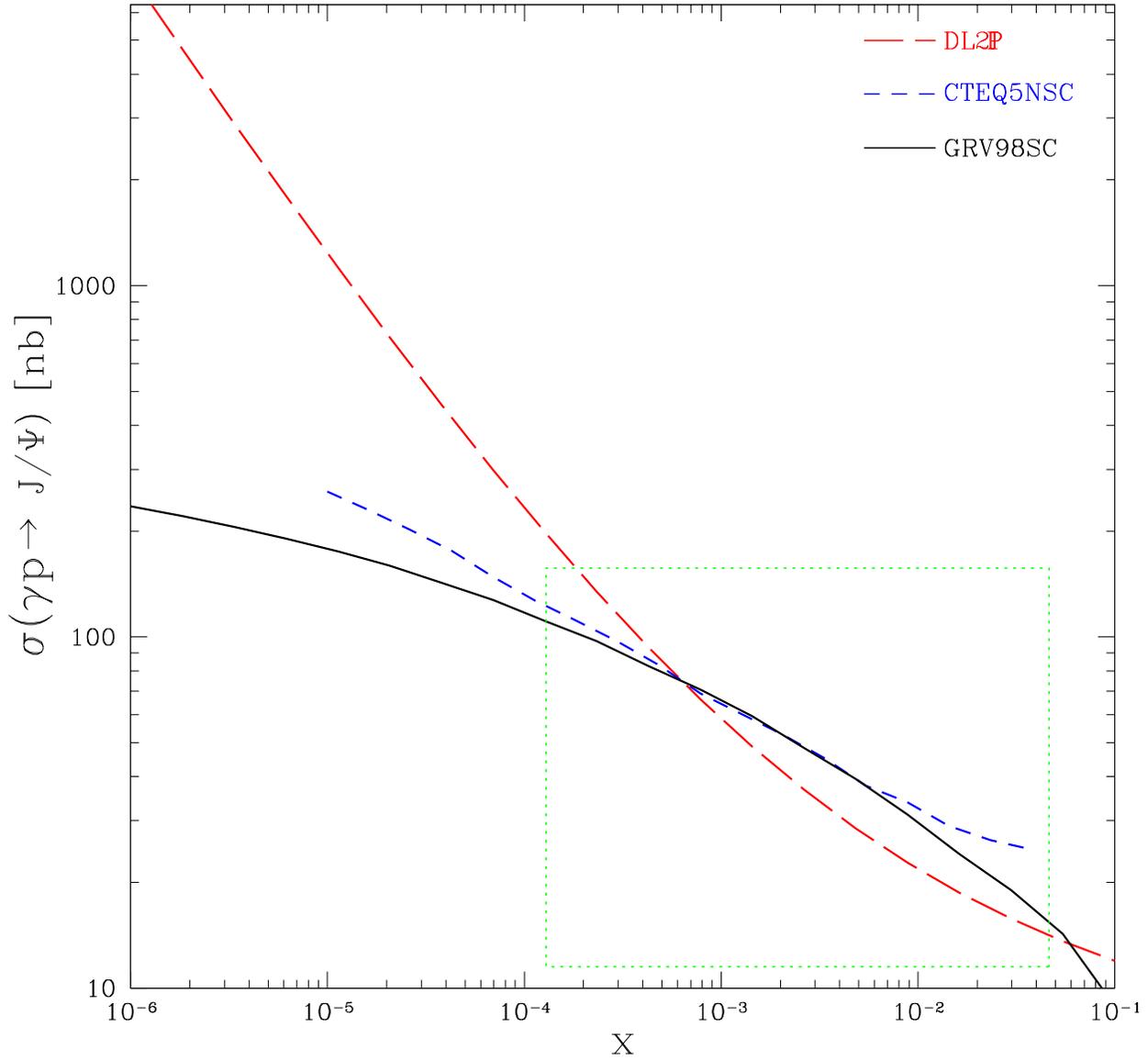,width=170mm}}
\caption{\it A comparison between
the energy dependence predictions of GRV98SC, CTEQ5NSC and DL2P 
for the integrated cross section of $J/\Psi$ photo production.
The available experimental data points are confined within the inner
window.}
\label{Fig.5}
\end{figure}
%%%%%%%%%%%%%%%%%%%%%%%%%%%%%%%%%%%%%%%%%%%%%%%%%%%%%%%%%%%%%%%%%%%%%%

\end{document}